\documentclass[amsmath,amssymb,aps,prd,twocolumn,10pt]{revtex4-2}

\usepackage{xcolor}
\usepackage{graphicx}
\usepackage{bm}
\usepackage[colorlinks=true, citecolor=jade, linkcolor=jade, urlcolor=jade]{hyperref}
\usepackage{ulem}

\definecolor{brightblue}{RGB}{0, 150, 255}
\definecolor{jade}{RGB}{0, 163, 108}

\begin{document}

\title{Low Primordial Helium as a Probe of Non-Unitary Neutrino Mixing}

\author{Shu-Yuan Guo}
\email{shyuanguo@ytu.edu.cn}
\author{Jia-Qi Li}%
\affiliation{Department of Physics, Yantai University, Yantai 264005, China}%

\begin{abstract}
The EMPRESS collaboration has reported a primordial helium mass fraction
\(Y_P = 0.2402 \pm 0.0040\), which lies more than \(1\sigma\) below the
standard Big Bang Nucleosynthesis prediction. Combined with the deuterium
abundance and the Planck baryon density, this corresponds to a mild
\(\sim 2\sigma\) tension with the standard value of the effective number of
relativistic species, or to a \(\sim 2\sigma\) indication of a non-zero
electron-neutrino degeneracy. We show that it can be explained by non-unitary
three-neutrino mixing, a generic feature of seesaw extensions of the Standard
Model. Non-unitarity makes the Fermi constant in beta decay exceed that in muon
decay, enhancing the neutron-to-proton conversion rate and lowering \(Y_P\). A
full numerical BBN calculation, combined with the EMPRESS \(Y_P\)
determination and the deuterium abundance  gives the best fit point at \(\alpha_{22}=0.982\), while the standard values
\(N_{\mathrm{eff}}=3.044\), \(\xi_e=0\) remain allowed; the required level of non-unitarity is consistent with current experimental constraints.
\end{abstract}

\maketitle

\section{Introduction}
\label{sec:intro}

Big Bang Nucleosynthesis (BBN) provides a unique probe of the first few
minutes of the Universe, when the primordial plasma cooled sufficiently
for the synthesis of the light elements. The resulting abundances are
exquisitely sensitive to the baryon-to-photon ratio \(\eta\), the expansion
rate, and the weak interaction rates that interconvert protons and
neutrons. In the standard cosmological model (\(\Lambda\)CDM) with three
flavors of light neutrinos, the predicted \(Y_P\) is essentially fixed by
the expansion rate around neutron freeze-out, conventionally parametrized
by the effective number of relativistic degrees of freedom \(N_{\rm eff}\),
while \(\eta\) enters only weakly and is pinned down by the cosmic
microwave background (CMB) at the percent level. The SM prediction for the effective number of relativistic species
is \(N_{\rm eff}=3.044\)~\cite{Froustey:2020mcq,Bennett:2020zkv}, the Planck determination of the baryon density is
\(\eta_{10}\equiv10^{10}\eta=6.132\pm0.038\)~\cite{Planck:2018vyg}, and
standard BBN yields a helium mass fraction \(Y_P=0.2470\pm0.0002\)
\cite{ParticleDataGroup:2024cfk}.

Over the past two decades, \(Y_P\) has been measured from the spectra of
metal-poor H II regions in dwarf galaxies. Earlier compilations give
\(Y_P \simeq 0.244\)--\(0.255\), broadly consistent with the standard
prediction within errors
\cite{Izotov:2014fga,Aver:2015iza,Hsyu:2020uqb,Kurichin:2021ppm,Valerdi:2019beb}.
Very recently, the EMPRESS collaboration has reported a determination
that lies systematically below this range. Based on 29 newly observed
galaxies (14 of them extremely metal-poor galaxies, EMPGs) combined with
58 from previous studies, it gives \(Y_P = 0.2402 \pm 0.0040\) (hereafter
EMPRESS 2025) \cite{Yanagisawa:2025mgx}. This supersedes the earlier EMPRESS analysis of a smaller sample, which gave the consistent value \(Y_P = 0.2370 \pm 0.0034\) \cite{Matsumoto:2022tlr}. The analysis
performs a linear regression over a broad metallicity range, including
both EMPGs and moderately metal-poor ones, and extrapolates the
He/H--O/H relation to zero metallicity. The $Y_P$ value lies more than \(1\sigma\) below the standard BBN prediction.
Combined with the deuterium abundance D/H~\cite{Cooke:2017cwo}, it corresponds
to a mild \(\sim 2\sigma\) tension with the standard value of
\(N_{\mathrm{eff}}\) and the Planck baryon density, or, equivalently, to a
\(\sim 2\sigma\) indication of a non-zero lepton asymmetry~\cite{Yanagisawa:2025mgx}.
It is important to note that the EMPRESS result is not isolated. The
ACT DR6 analysis reports a direct CMB constraint
\(Y_P = 0.227 \pm 0.014\) \cite{Calabrese:2025}. This central value lies
below the standard BBN prediction, in the same direction as the EMPRESS
determination, although its precision does not yet allow a firm conclusion.

On the other hand, the LBT \(Y_p\) project---the most precise determination to date---has recently released its measurement result, \(Y_p = 0.2458 \pm 0.0013\), obtained by averaging its 15 lowest-metallicity targets (\(\mathrm{O/H} < 4 \times 10^{-5}\))~\cite{Aver:2026dxv}.
This result agrees with the standard BBN
prediction. The divergence from the EMPRESS central value is largely one
of method: the LBT analysis restricts the sample to its lowest-metallicity
targets and takes their weighted mean, whereas the EMPRESS analysis fits
the He/H--O/H relation over a broader metallicity range and extrapolates it
to zero metallicity. Whether the residual difference reflects a genuine
evolutionary slope, systematics affecting the EMPG sample, or a real shift
of the primordial abundance remains an open question.

Rather than claiming that the EMPRESS result
resolves the tension in favor of new physics, we take its central value as
a well-motivated benchmark that current data still allow, and explore its
cosmological consequences. The two standard handles on low \(Y_P\) are a
modified \(N_{\rm eff}\), which changes the expansion rate at freeze-out,
and a non-zero electron-neutrino degeneracy \(\xi_e\), which shifts the
beta equilibrium \(n+\nu_e \leftrightarrow p+e^-\) toward a lower
neutron-to-proton ratio. The EMPRESS collaboration itself explored both
possibilities: its latest analysis finds \(N_{\rm eff}=2.54^{+0.20}_{-0.25}\)
for \(\xi_e=0\), while allowing a non-zero degeneracy gives
\(\xi_e=0.05^{+0.02}_{-0.03}\)~\cite{Yanagisawa:2025mgx}. Other proposals
invoke early dark energy~\cite{Takahashi:2022cpn,McKeen:2024voa}, a varying
fine-structure constant~\cite{Seto:2023yal} or Higgs vacuum expectation
value~\cite{Meyer:2024auq}, or CPT
violation~\cite{Barenboim:2026ocd}.

A particularly well-motivated alternative is to modify the weak interaction rates that govern neutron
freeze-out, as arises naturally when the effective low-energy neutrino
mixing matrix is non-unitary because of heavy sterile states. In the early
Universe, non-unitarity modifies both neutrino decoupling and, more
importantly, the effective Fermi constant entering \(\beta\) decay. Since
\(G_F^\beta\) can exceed the muon-decay Fermi constant \(G_F^\mu\), the
neutron-to-proton conversion rate is enhanced, reducing the neutron
abundance at freeze-out and hence \(Y_P\). In this work we show that a
small deviation from unitarity, accounts for the low \(Y_P\) reported by
EMPRESS, is consistent with the measured D/H and with the CMB
determination of \(\eta\), and requires no modification of \(N_{\rm eff}\)
or \(\xi_e\).

The remainder of this paper is organized as follows. In Sec.~\ref{sec:fw},
we present the formalism of non-unitary neutrino mixing and its
implementation into the weak interaction rates. Sec.~\ref{sec:res} reports
our numerical results, together with the combined constraints from BBN and
terrestrial experiments. Finally, Sec.~\ref{sec:conc} summarizes our
findings and discusses their implications for probing new physics with
primordial nucleosynthesis.

\section{Non-Unitary Mixing and BBN}
\label{sec:fw}

Non-unitary three-neutrino mixing arises in extensions of the SM that generate neutrino masses via additional sterile states. When three light neutrinos mix with \(n-3\) heavy states, the full \(n\times n\) lepton mixing matrix \(K\) is unitary and can be decomposed as~\cite{Hettmansperger:2011bt}
\begin{equation}
K = \begin{pmatrix} N_{3\times 3} & S_{3\times (n-3)} \\ V_{(n-3)\times 3} & T_{(n-3)\times (n-3)} \end{pmatrix},
\label{eq:Kdecomp}
\end{equation}
where \(N\) describes mixing among the three light neutrinos, \(S\) and \(V\) encode light--heavy mixing, and \(T\) describes heavy--heavy mixing. Although \(K\) is unitary, the light sub-block \(N\) need not be. Deviations from unitarity are parameterized as~\cite{Escrihuela:2015wra}
\begin{equation}
N = \begin{pmatrix} \alpha_{11} & 0 & 0 \\ \alpha_{21} & \alpha_{22} & 0 \\ \alpha_{31} & \alpha_{32} & \alpha_{33} \end{pmatrix} U_{\mathrm{PMNS}},
\label{eq:Nparam}
\end{equation}
where \(U_{\mathrm{PMNS}}\) is the standard unitary PMNS matrix, the diagonal \(\alpha_{ii}\)s are real, and the off-diagonal \(\alpha_{ij}\)s may be complex. In the SM, \(\alpha_{ij}=\delta_{ij}\).

The existence of non-zero \(\alpha_{ij}\) modifies neutrino weak interactions and leads to a mismatch between the Fermi constant measured in muon decay, \(G_F^\mu\), and the one relevant for beta decay, \(G_F^\beta\):
\begin{eqnarray}
G_F^\mu &=& G_F \sqrt{(N N^\dagger)_{11}(N N^\dagger)_{22}}, \\
G_F^\beta &=& G_F \sqrt{(N N^\dagger)_{11}} = \frac{G_F^\mu}{\sqrt{(N N^\dagger)_{22}}},
\label{eq:2.1b}
\end{eqnarray}
with \(G_F\) the Fermi constant in the SM. This mismatch is the key to understanding the impact of non-unitary mixing on BBN.

The light-element abundances are primarily determined by the neutron-to-proton ratio at the onset of nucleosynthesis. The relevant weak processes that interconvert neutrons and protons are \(n+\nu_e\leftrightarrow p+e^-\), \(n+e^+\leftrightarrow p+\bar\nu_e\), and \(n\leftrightarrow p+e^-+\bar\nu_e\). Their rates are proportional to
\begin{equation}
\Gamma_{n\leftrightarrow p}\propto (G_F^\beta)^2 V_{ud}^2(1+3g_A^2),
\label{eq:weakrate}
\end{equation}
where \(V_{ud}\) is the CKM matrix element and \(g_A\) is the axial-vector coupling. In the presence of non-unitarity, \(G_F^\beta\) is generally larger than \(G_F^\mu\). This follows from the unitarity constraint, which implies
\begin{equation}
\sqrt{(N N^\dagger)_{22}} = \sqrt{\alpha_{22}^2+|\alpha_{21}|^2}\le 1,
\label{eq:unitaritybound}
\end{equation}
so that
\begin{equation}
G_F^\beta = \frac{G_F^\mu}{\sqrt{\alpha_{22}^2+|\alpha_{21}|^2}} \ge G_F^\mu.
\label{eq:GFbeta_ge}
\end{equation}
A larger \(G_F^\beta\) enhances the neutron-to-proton conversion rates. As a result, the neutron-to-proton ratio at freeze-out decreases, and the resulting abundances of \(^4\mathrm{He}\) and deuterium are reduced. Note that $\alpha_{21}$ and 
$\alpha_{22}$ in Eq.~(\ref{eq:GFbeta_ge}) are partially degenerate to affect the weak rate. For simplicity, we set 
$\alpha_{21} = 0$ in the following and attribute the whole effect to 
$\alpha_{22}$.

Non-unitary mixing also affects neutrino decoupling and hence \(N_{\mathrm{eff}}\), defined as
\begin{equation}
N_{\mathrm{eff}}
=
\left(\frac{11}{4}\right)^{4/3}
\sum_\alpha
\left(\frac{T_{\nu_\alpha}}{T_\gamma}\right)^4,
\label{eq:Neffdef}
\end{equation}
where \(T_{\nu_\alpha}\) and \(T_\gamma\) are the neutrino and photon temperatures. The parameters \(\alpha_{ij}\) modify both neutral-current (NC) and charged-current (CC) neutrino interactions. In the low-energy regime, these effects can be mapped onto neutrino NC non-standard interactions with electrons, as done in the literature~\cite{Escrihuela:2015wra,Barenboim:2025okj}. This mapping changes the weak couplings and the energy-transfer rates, thereby altering the evolution of \(T_{\nu_\alpha}\) and the ratio \(T_\nu/T_\gamma\), and hence \(N_{\mathrm{eff}}\). A change in \(N_{\mathrm{eff}}\) modifies the radiation energy density \(\rho_R\) and the Hubble rate, \(H\propto\sqrt{\rho_R}\). Since freeze-out occurs when \(\Gamma_{\mathrm{weak}}(T_f)\sim H(T_f)\), this shifts the freeze-out temperature \(T_f\) and thus the neutron-to-proton ratio, providing an indirect BBN effect. In our full numerical calculations we implement this mapping following the procedure of Refs.~\cite{Escrihuela:2015wra,Barenboim:2025okj}.

For nuclear abundances, however, this \(N_{\mathrm{eff}}\) channel is subdominant to the direct modification of the beta-decay Fermi constant \(G_F^\beta\). The latter enters the \(n\leftrightarrow p\) rates as \(\Gamma_{n\leftrightarrow p}\propto (G_F^\beta)^2\), so a change in \(G_F^\beta\) directly rescales the weak rates. In contrast, \(N_{\mathrm{eff}}\) enters only through \(H\), and therefore affects \(T_f\) only indirectly and weakly. Since the \(N_{\mathrm{eff}}\) channel is subdominant, in the linear response analysis below we characterize
the strength of the non-unitary effect primarily through the direct shift
of the beta-decay Fermi constant, \(\Delta G_F^\beta \equiv G_F^\beta -
G_F^\mu\).

To describe the response of the light-element abundances we use the
response-matrix formalism of Ref.~\cite{Dent:2007zu}, in which the sensitivity
of an abundance \(Y_a\) to a parameter \(X_i\) is defined by
\begin{equation}
S_{a,i}=\frac{\partial\ln Y_a}{\partial\ln X_i},
\label{eq:cdent}
\end{equation}
evaluated with all other parameters held fixed. For the two
abundances and the two parameters relevant here, \(Y_a\in\{Y_P,\rm D/H\}\) and
\(X_i\in\{G_F^\beta,\eta\}\), we have
\begin{eqnarray}
\frac{\Delta Y_P}{Y_P}
&=&
S_{Y_P,G}\,\Delta\ln G_F^\beta
+
S_{Y_P,\eta}\frac{\Delta\eta}{\eta},
\label{eq:linYP}
\\
\frac{\Delta(\rm D/H)}{\rm D/H}
&=&
S_{D,G}\,\Delta\ln G_F^\beta
+
S_{D,\eta}\frac{\Delta\eta}{\eta}.
\label{eq:linD}
\end{eqnarray}
Here \(\Delta \ln G_F^\beta \equiv \ln(G_F^\beta/G_F^\mu)\) denotes the
logarithmic shift of the beta-decay Fermi constant relative to the
muon-decay value $G_F^\mu$, so that \(\Delta \ln G_F^\beta \approx \Delta G_F^\beta /
G_F^\mu\) for small deviations. 
We keep \(\eta\) free because it is the only standard BBN parameter that
can reduce \(Y_P\) without invoking new physics; fixing it a priori would
therefore presume the conclusion. As we show below, however, its
influence on \(Y_P\) is far too small to account for the EMPRESS central
value. The deuterium abundance then serves two purposes: it determines
\(\eta\) at the per cent level independently of the CMB, and it responds
to the same \(G_F^\beta\), so that it provides a consistency test of the
non-unitary scenario rather than merely calibrating \(\eta\).
The response to the baryon density is then read directly from the response
matrix of Ref.~\cite{Dent:2007zu}, \(S_{Y_P,\eta}\simeq 0.04\) and
\(S_{D,\eta}\simeq -1.6\), the latter reflecting the familiar scaling
\(\rm D/H\propto\eta^{-1.6}\). The weak-rate coefficients follow from the same
matrix through the neutron lifetime. Because \(\tau_n\) enters only the
normalisation of the \(n\leftrightarrow p\) rates and of the free neutron
decay, a variation of \(G_F^\beta\) is equivalent to a variation of
\(\tau_n\)~\cite{Dent:2007zu}, with the one-to-one mapping
\begin{equation}
\Delta\ln\tau_n=-2\,\Delta\ln G_F^\beta,
\label{eq:taumap}
\end{equation}
which simply reflects \(\tau_n\propto(G_F^\beta)^{-2}\). The \(\tau_n\) entries
of the response matrix, \(\partial\ln Y_P/\partial\ln\tau_n\simeq 0.73\) and
\(\partial\ln(\rm D/H)/\partial\ln\tau_n\simeq 0.41\)~\cite{Dent:2007zu}, therefore give
\begin{equation}
S_{Y_P,G}\simeq-1.5,
\qquad
S_{D,G}\simeq-0.8 .
\label{eq:SYPG}
\end{equation}
Using \(\Delta\ln G_F^\beta=-\ln\alpha_{22}\), the linear
response to the non-unitary parameter becomes
\begin{eqnarray}
\frac{\Delta Y_P}{Y_P}
&\simeq&
1.5\,\Delta\alpha_{22}
+
0.04\frac{\Delta\eta}{\eta},
\label{eq:approxYP}
\\
\frac{\Delta(\rm D/H)}{\rm D/H}
&\simeq&
0.8\,\Delta\alpha_{22}
-
1.6\frac{\Delta\eta}{\eta}.
\label{eq:approxD}
\end{eqnarray}
Here \(\Delta\alpha_{22}\equiv\alpha_{22}-1\) denotes the deviation of $\alpha_{22}$ from its standard value. These coefficients provide a useful qualitative guide to the numerical
analysis, while the quantitative constraints presented in
Sec.~\ref{sec:res} are obtained from the full BBN network.

\section{Results and Discussions}
\label{sec:res}

\begin{figure}[htbp]
    \centering
    \includegraphics[width=\columnwidth]{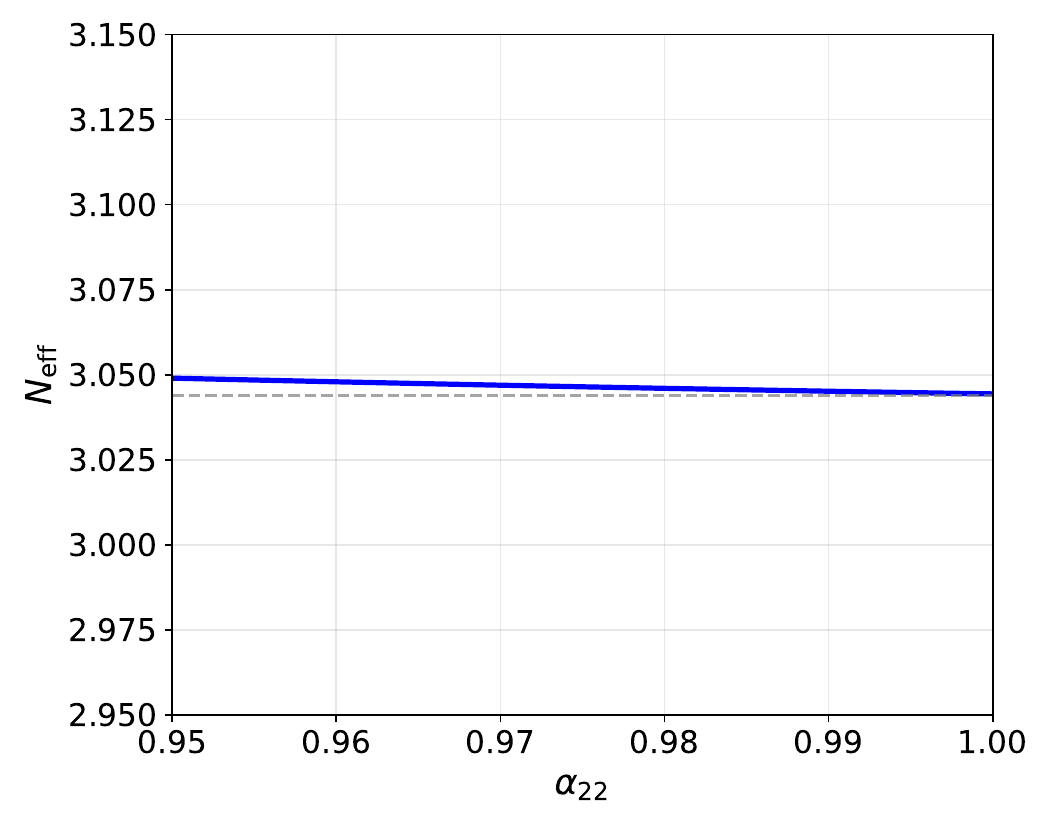}
    \caption{
        Effective number of relativistic species \(N_{\mathrm{eff}}\) induced by
        non-unitary mixing, as a function of \(\alpha_{22}\). The gray dashed line indicates the SM value $3.044$.
    }
    \label{fig:Neff_alpha22}
\end{figure}

Having established the analytic scaling relations in Sec.~\ref{sec:fw}, we now
turn to the full numerical computation of the primordial abundances in the
presence of non-unitary mixing. The abundances are computed with
\texttt{PRyMordial}~\cite{Burns:2023pry}, a public Python package for precision
BBN calculations that returns the light-element abundances together with \(N_{\mathrm{eff}}\), starting from a self-consistent
treatment of the thermal background---including non-instantaneous neutrino
decoupling---and using ab-initio \(n\leftrightarrow p\) weak rates with radiative
and finite-temperature corrections. For the key thermonuclear reactions we adopt
the NACRE~II compilation~\cite{Xu:2013fha}. As discussed above, non-unitary
mixing affects BBN through two channels: the direct modification of the
beta-decay Fermi constant \(G_F^\beta\), which changes the \(n\leftrightarrow p\)
weak rates, and the indirect modification of neutrino decoupling, which alters
\(N_{\mathrm{eff}}\) and hence the Hubble expansion rate. In our numerical BBN
calculation, both effects are implemented self-consistently: for each value of
\(\alpha_{22}\), the corresponding \(N_{\mathrm{eff}}\) is computed from the
non-unitary mapping to neutrino NC non-standard interactions described in Sec.~\ref{sec:fw}, and the
weak rates are modified according to \(G_F^\beta=G_F^\mu/\alpha_{22}\). The indirect \(N_{\mathrm{eff}}\) effect is numerically negligible.
Figure~\ref{fig:Neff_alpha22} shows the value of \(N_{\mathrm{eff}}\) induced by
non-unitary mixing as a function of \(\alpha_{22}\): it stays within
$6\times 10^{-3}$  of the SM value \(3.044\) over the range
of interest, and the corresponding change in the abundances is far below the
observational precision.

\begin{figure*}[htbp]
    \centering
    \includegraphics[width=2\columnwidth]{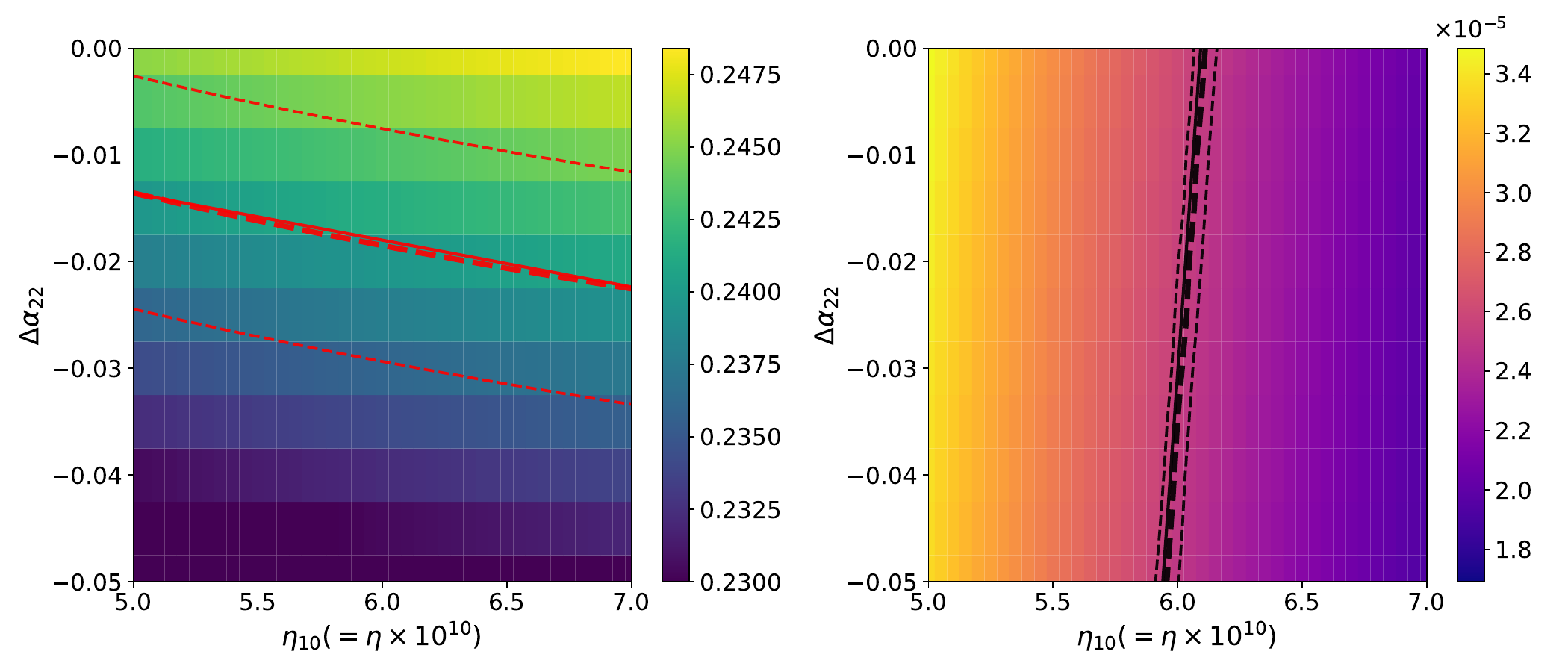}
    \caption{
        The primordial helium mass fraction $Y_P$ (left) and the deuterium abundance $\mathrm{D/H}$ (right) in the $\eta_{10}$--$\Delta\alpha_{22}$ plane. The thick and thin dashed lines indicate the adopted central values and \(1\sigma\) intervals: \(Y_P=0.2402\pm0.0040\), \(\mathrm{D/H}=(2.527\pm0.030)\times10^{-5}\). The solid lines show the predictions of Eqs.~\ref{eq:approxYP} and~\ref{eq:approxD} for the central values of \(Y_P\) and \(\mathrm{D/H}\).}
    \label{fig:alpha22_eta}
\end{figure*}

Figure~\ref{fig:alpha22_eta} displays the resulting theoretical predictions for
\(Y_P\) (left panel) and \(\mathrm{D/H}\) (right panel) in the
\(\eta_{10}\)--\(\Delta \alpha_{22}\) plane. In each panel the thick dashed
line marks the measured central value, \(Y_P=0.2402\) on the left and
\(\mathrm{D/H}=2.527\times10^{-5}\)~\cite{Cooke:2017cwo} on the right, the thin dashed lines delimit
the corresponding \(1\sigma\) intervals,  and the solid line is the locus
on which the analytic relation of Eq.~\eqref{eq:approxYP}
or~\eqref{eq:approxD} reproduces the same central value. The left panel shows
that \(Y_P\) depends strongly on \(\alpha_{22}\): over the range shown, \(Y_P\)
falls from \(0.2470\) at \(\Delta\alpha_{22}=0\) to about \(0.23\) at
\(\Delta\alpha_{22}=-0.05\). The dependence on \(\eta\) is much weaker,
reflecting the small positive sensitivity of \(Y_P\) to the baryon density. In
the right panel, \(\mathrm{D/H}\) is primarily controlled by \(\eta\),
decreasing as \(\eta\) increases, while its dependence on \(\alpha_{22}\) is
weaker than that of \(Y_P\) but not negligible, which tilts the \(1\sigma\)
band in the \(\alpha_{22}\)--\(\eta_{10}\) plane. The different slopes of the
two sets of contours imply that \(Y_P\) and \(\mathrm{D/H}\) constrain
complementary directions in the parameter space, allowing \(\alpha_{22}\) and
\(\eta\) to be determined simultaneously. In both panels the solid line
follows the thick dashed one, this confirms that the linear response analysis of
Sec.~\ref{sec:fw} reproduces the behavior of the full network here.

\begin{figure}[htbp]
    \centering
    \includegraphics[width=\columnwidth]{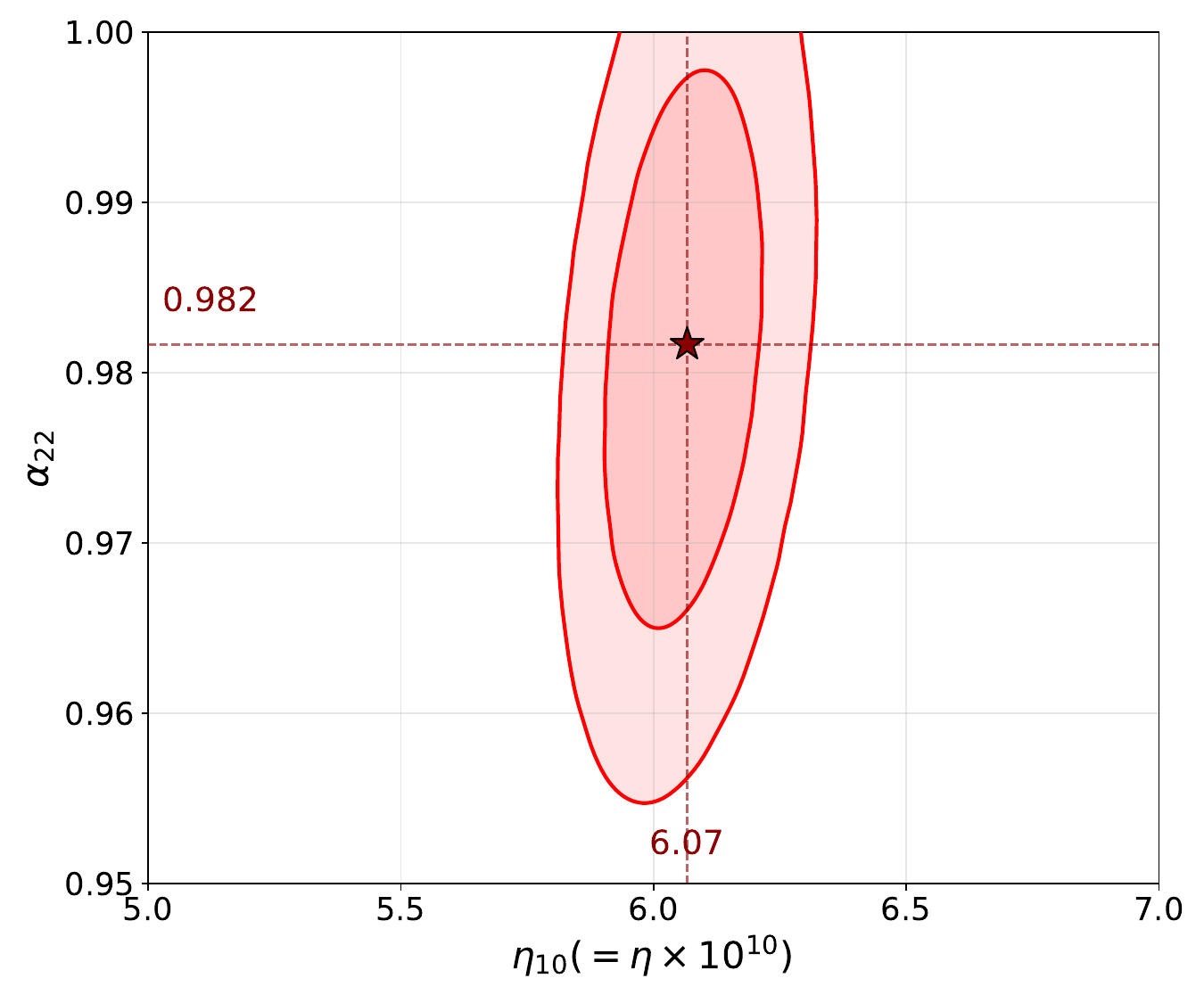}
    \caption{
        The $1\sigma$ and $2\sigma$ allowed regions in the $\eta_{10}$--$\alpha_{22}$ plane from the combined EMPRESS 2025 measurements of $Y_P$ and $\mathrm{D/H}$. The red shaded areas denote the confidence regions, and the star marks the best-fit point at $\alpha_{22}=0.982$ and $\eta_{10}=6.07$.
    }
    \label{fig:contour_alpha22_eta}
\end{figure}

Figure~\ref{fig:contour_alpha22_eta} shows the \(1\sigma\) and \(2\sigma\)
constraints in the $\eta_{10}$--$\alpha_{22}$ plane obtained from the
EMPRESS 2025 measurements of \(Y_P\) and \(\mathrm{D/H}\). The SM value
\(\alpha_{22}=1\) lies outside the \(1\sigma\) region, and the best-fit
point is at \(\alpha_{22}=0.982\) and \(\eta_{10}=6.07\).
Non-unitary mixing is constrained by terrestrial experiments; here we adopt
the limits obtained from neutrino oscillation data, \(\alpha_{22}>0.98\) at
\(3\sigma\)~\cite{Barenboim:2025okj,Forero:2021azc}. The unitarity test of the first row of
the CKM matrix, combined with the oscillation bound on \(\alpha_{21}\),
would imply the much stronger limit \(\alpha_{22}>0.999\) at \(90\%\)
C.L.~\cite{Escrihuela:2015wra}. Since such a test assumes no new physics in
the quark sector, whereas the mild tensions presently seen in CKM
unitarity tests might themselves have a new-physics
origin~\cite{Cirigliano:2022yyo}, we adopt the more conservative
oscillation-based constraint.
Thus the non-unitary framework provides a viable
explanation of the EMPRESS result, the required
deviation lies just inside the oscillation-based bound. The low \(Y_P\) is therefore
reproduced by a single new-physics parameter.

Although the estimates presented above focus on the impact of the non-unitary parameter \(\alpha_{22}\) and the baryon-to-photon ratio \(\eta\) on the light-element abundances, the observed low values of \(Y_P\) reported by EMPRESS have been interpreted in the literature in terms of other cosmological parameters. In particular, the EMPRESS collaboration itself has explored explanations based on a modified effective number of relativistic species, \(N_{\mathrm{eff}}\), and on a non-zero electron-neutrino degeneracy parameter, \(\xi_e\equiv \mu_{\nu_e}/T_\nu\). A lower \(N_{\mathrm{eff}}\) changes the Hubble expansion rate and shifts the neutron-proton freeze-out temperature, while a positive \(\xi_e\) shifts the beta equilibrium \(n+\nu_e\leftrightarrow p+e^-\) toward a lower neutron-to-proton ratio; both effects can reduce \(Y_P\) and therefore partially mimic the influence of \(\alpha_{22}\).
To assess whether non-unitary mixing can provide a more economical or more consistent explanation of the data, we therefore include \(N_{\mathrm{eff}}\) and \(\xi_e\) as free parameters in our statistical analysis and perform a joint fit to the primordial abundances. It should be emphasised that the \(N_{\mathrm{eff}}\) varied here is not the shift induced by non-unitary mixing, which is negligible (Fig.~\ref{fig:Neff_alpha22}), but the contribution of any additional source of relativistic degrees of freedom beyond those already accounted for in the above treatment. It is introduced solely as a free parameter for this statistical comparison, and we do not discuss the details of its possible origin.

\begin{figure}[htbp]
    \centering
    \includegraphics[width=\columnwidth]{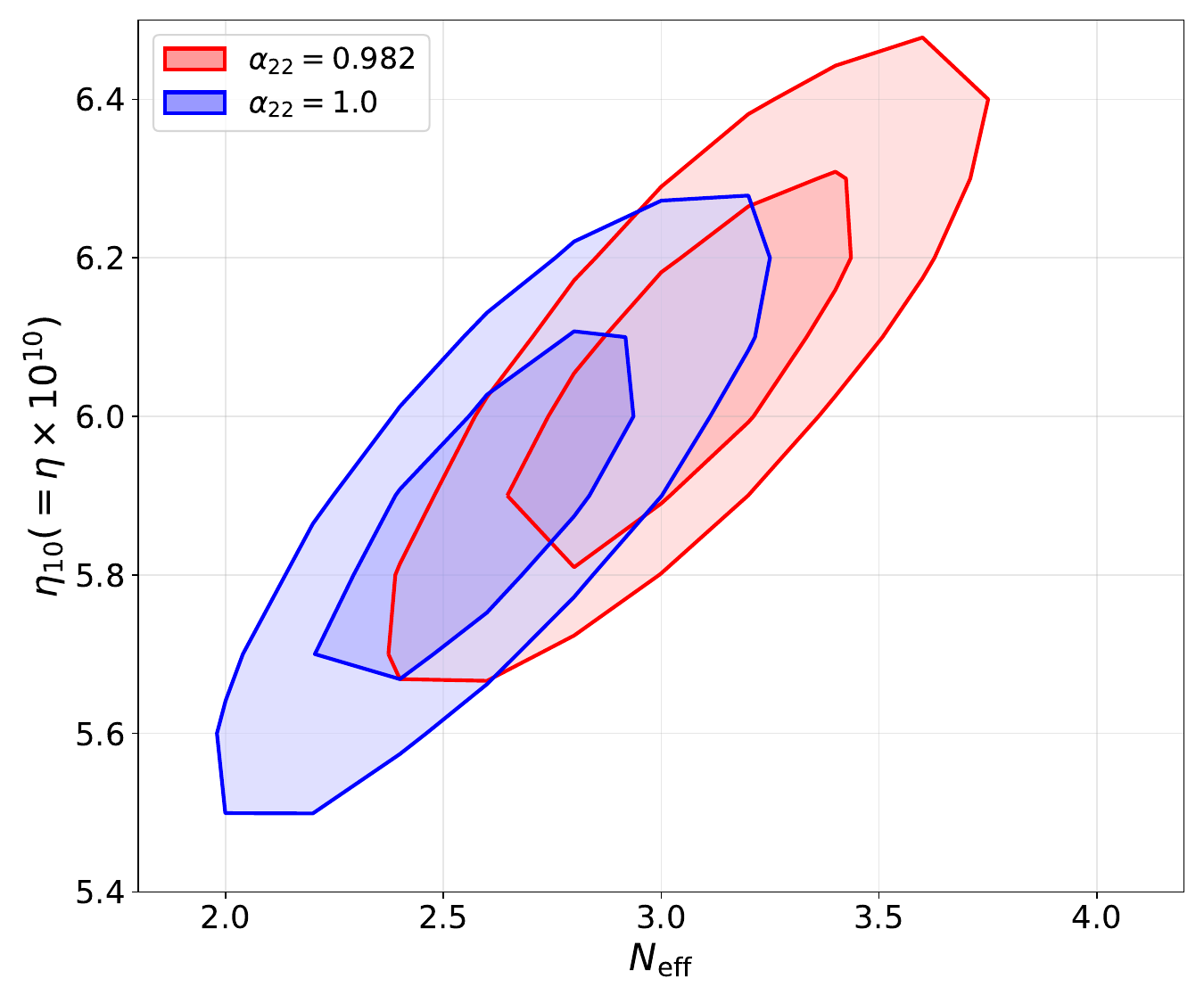}
    \caption{
        The $1\sigma$ and $2\sigma$ constraints in the $N_{\mathrm{eff}}$--$\eta_{10}$ plane from the EMPRESS 2025 measurements of $Y_P$ and $\mathrm{D/H}$, with $\xi_e=0$. The red shaded region corresponds to $\alpha_{22}=0.982$, and the blue contours to $\alpha_{22}=1.0$.
    }
    \label{fig:contour_Neff_eta}
\end{figure}

The total chi-square function is defined as
\begin{equation}
\chi^2(\alpha_{22},\eta,\xi_e,N_{\mathrm{eff}})
=
\chi^2_{Y_P}
+
\chi^2_{\rm D/H}
+
\chi^2_{\eta},
\label{eq:chi2_total}
\end{equation}
where the individual terms are given by
\begin{align}
\chi^2_{Y_P}
&=
\frac{
\left[
Y_P^{\mathrm{obs}}
-
Y_P^{\mathrm{th}}(\alpha_{22},\eta,\xi_e,N_{\mathrm{eff}})
\right]^2
}{
\sigma_{Y_P,\mathrm{obs}}^2
+
\sigma_{Y_P,\mathrm{th}}^2
},
\label{eq:chi2_YP}
\\[4pt]
\chi^2_{\rm D/H}
&=
\frac{
\left[
(\rm D/H)^{\mathrm{obs}}
-
(\rm D/H)^{\mathrm{th}}(\alpha_{22},\eta,\xi_e,N_{\mathrm{eff}})
\right]^2
}{
\sigma_{\rm D/H,\mathrm{obs}}^2
+
\sigma_{\rm D/H,\mathrm{th}}^2
},
\label{eq:chi2_DH}
\\[4pt]
\chi^2_{\eta}
&=
\frac{
\left(
\eta_{\mathrm{Planck}}
-
\eta
\right)^2
}{
\sigma_{\eta,{\mathrm{Planck}}}^2
}.
\label{eq:chi2_eta}
\end{align}
The first two terms compare the observed abundances with the theoretical predictions of our modified BBN calculation, incorporating both observational and theoretical uncertainties; for the latter we adopt \(\sigma_{Y_P,\mathrm{th}}\simeq1.2\times10^{-4}\), dominated by the nuclear reaction rates and the neutron lifetime, and \(\sigma_{\rm D/H,\mathrm{th}}\simeq0.06\times10^{-5}\)~\cite{Seto:2023yal}. The third term is an external Gaussian prior on \(\eta\) from the Planck measurement of the baryon density.
When presenting two-dimensional constraints in a plane that involves \(\eta\), we omit \(\chi^2_{\eta}\) from the total chi-square. For one-dimensional constraints, and for planes that do not involve \(\eta\) (such as \(N_{\mathrm{eff}}\)--\(\xi_e\)), the Planck prior is always included. In all two-dimensional contours, we profile over the remaining parameters by minimizing the chi-square with respect to them.

Before presenting the combined constraints on \(N_{\mathrm{eff}}\) and \(\xi_e\), it is instructive to examine the parameter space preferred by the $N_{\rm eff}$-only interpretation of the low \(Y_P\). Figure~\ref{fig:contour_Neff_eta} shows the \(1\sigma\) and \(2\sigma\) allowed regions in the \(N_{\mathrm{eff}}\)--\(\eta_{10}\) plane obtained from the EMPRESS 2025 measurements combined with the deuterium abundance. The blue contours correspond to \(\alpha_{22}=1.0\), while the red shaded regions correspond to the non-unitary scenario with \(\alpha_{22}=0.982\). For \(\alpha_{22}=1.0\), the data show a clear preference for lower values of both \(N_{\mathrm{eff}}\) and \(\eta\): the \(1\sigma\) region extends down to \(N_{\mathrm{eff}}\simeq 2.2\) and \(\eta_{10}\simeq 5.7\), with the best fit point at \(N_{\mathrm{eff}}\simeq 2.54\) and \(\eta_{10}\simeq 5.88\). This tendency is driven by the need to reduce the predicted \(Y_P\): since \(Y_P\) increases with \(N_{\mathrm{eff}}\), a smaller \(N_{\mathrm{eff}}\) lowers the helium abundance, while a smaller \(\eta\) also slightly reduces \(Y_P\) and helps to accommodate the low central value reported by EMPRESS. This is the same tendency that the EMPRESS collaboration itself reports~\cite{Yanagisawa:2025mgx}. This region is, however, difficult to accommodate. The standard point
\(N_{\mathrm{eff}}=3.044\), \(\eta_{10}=6.132\) lies outside the
\(1\sigma\) contour. For \(\alpha_{22}=0.982\), by contrast, the standard point
lies well inside the \(1\sigma\) contour, the best fit point is at
\(N_{\mathrm{eff}}=3.05\) and
\(\eta_{10}=6.08\). Non-unitary mixing therefore
reproduces the low \(Y_P\) with the cosmological parameters left at their
standard values, whereas the SM interpretation requires a departure from them.

\begin{figure}[htbp]
    \centering
    \includegraphics[width=\columnwidth]{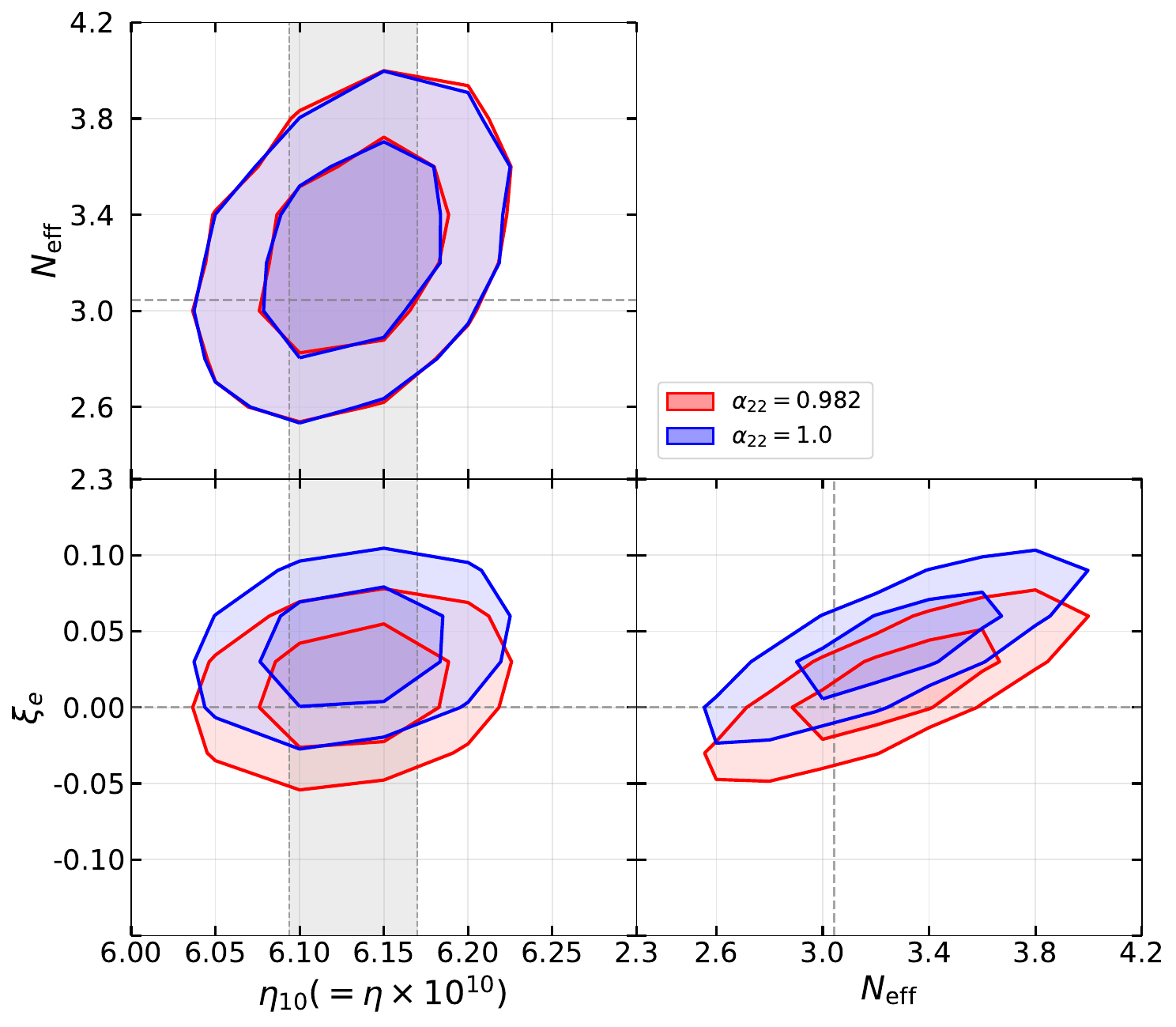}
    \caption{
        The $1\sigma$ and $2\sigma$ constraints projected onto the $\eta_{10}$--$N_{\mathrm{eff}}$ (upper), $\eta_{10}-\xi_e$ (lower left) and $N_{\mathrm{eff}}$--$\xi_e$ (lower right) planes for $\alpha_{22}=0.982$ (red shaded) and $\alpha_{22}=1.0$ (blue shaded), with $N_{\mathrm{eff}}$, $\xi_e$, and $\eta_{10}$ varied. Whereas Fig.~\ref{fig:contour_Neff_eta} assumes a vanishing lepton asymmetry, here $\xi_e$ is left free. Gray dashed lines indicate the standard values $N_{\mathrm{eff}}=3.044$, and $\xi_e=0$; the gray band represents the Planck $1\sigma$ constraint on $\eta$.
    }
    \label{fig:contour_xi_eta}
\end{figure}

The EMPRESS collaboration has suggested that the same data may instead be
described with the radiation density at its standard value and a non-zero
lepton asymmetry~\cite{Yanagisawa:2025mgx}. We examine this possibility in
Fig.~\ref{fig:contour_xi_eta}. It shows the
\(1\sigma\) and \(2\sigma\) constraints obtained from the same data once
\(\xi_e\) is allowed to vary, projected onto the
\(\eta_{10}\)--\(N_{\mathrm{eff}}\) plane (upper panel) and onto \(\xi_e\)
(lower panels). Two features are worth emphasising.
First, freeing \(\xi_e\) removes the preference for a low \(N_{\mathrm{eff}}\).
In the upper panel the standard value \(N_{\mathrm{eff}}=3.044\) lies well
inside the \(1\sigma\) region for both sets of contours, and the contours
overlap the Planck range of \(\eta\); the pronounced downward shift found in
Fig.~\ref{fig:contour_Neff_eta} for \(\alpha_{22}=1.0\) is absent, because the
same reduction of \(Y_P\) can now be obtained from a positive lepton
asymmetry.
Second, the two scenarios differ in the status of the standard point
\(\xi_e=0\).  For \(\alpha_{22}=1.0\), the \(1\sigma\) region
instead lies at positive values, its lower edge lying close to
\(\xi_e=0\), the best fit point is at $\xi_e \simeq 0.05$. For \(\alpha_{22}=0.982\), the \(1\sigma\) region readily
contains \(\xi_e=0\), spanning \(\xi_e\simeq -0.02\) to
\(0.06\) and \(N_{\mathrm{eff}}\simeq 2.9\)
to \(3.7\): the standard cosmological parameters remain
allowed without any adjustment.
Taken together, Figs.~\ref{fig:contour_Neff_eta}
and~\ref{fig:contour_xi_eta} show that within the non-unitary framework the low
\(Y_P\) reported by EMPRESS is reproduced with neither a deviation in
\(N_{\mathrm{eff}}\) nor a deviation in \(\xi_e\): a single parameter
\(\alpha_{22}\), still compatible with the laboratory limits on the
leptonic mixing matrix, is sufficient.

\section{Conclusion}
\label{sec:conc}

We have investigated whether the low primordial helium abundance \(Y_P\)
reported by EMPRESS can be explained by non-unitary three-neutrino mixing.
Non-unitarity makes the beta-decay Fermi constant exceed the muon-decay one,
\(G_F^\beta>G_F^\mu\), enhancing the neutron-to-proton conversion rate and
thereby lowering \(Y_P\); the effect is dominated by the single parameter
\(\alpha_{22}\), the induced change in \(N_{\mathrm{eff}}\) being negligible.
Using a full numerical BBN calculation based on \texttt{PRyMordial}, together
with the EMPRESS 2025 \(Y_P\) measurement and the
deuterium abundance, we find the best fit
\(\alpha_{22}=0.982\).
We further compared the non-unitary scenario with the modified $N_{\rm eff}$ or lepton asymmetry scenario considered by the EMPRESS collaboration. Both alternatives require a departure of the cosmological parameters from their
standard values, whereas the non-unitary scenario does not. For
\(\alpha_{22}=1.0\), the fit with \(\xi_e=0\) pulls
\(N_{\mathrm{eff}}\) to \(2.54\), to be compared with the
standard value \(3.044\); allowing a non-zero
\(\xi_e\) restores \(N_{\mathrm{eff}}\) to its standard value but moves the fit
to \(\xi_e\simeq 0.05\). Non-unitary
mixing instead reproduces the low \(Y_P\) with \(N_{\mathrm{eff}}\), \(\xi_e\)
and \(\eta\) left at their standard values, the required value of $\alpha_{22}$ can be accommodated within the oscillation-based constraint.
Improved determinations of \(Y_P\) and D/H, together with forthcoming
terrestrial neutrino experiments, will test this scenario further and help to
clarify the origin of the low \(Y_P\).

\begin{acknowledgments}
This work by S.-Y. G. was supported by the NNSFC under Grant No. 12305113, and by the Natural Science Foundation
of Shandong Province under Grants No. ZR2022QA026.
\end{acknowledgments}



\nocite{*}

\bibliography{apssamp}

\end{document}